\documentclass{PoS}
\renewcommand{\ij}{i\!j}

\newcommand{\as}{\alpha_{\tiny{\mbox{S}}}}
\newcommand{\eps}{\epsilon}

\newcommand{\sig}{\sigma}

\newcommand{\npo}{{n\!+\!1}}
\newcommand{\npt}{{n\!+\!2}}

\newcommand{\mc}{\mathcal}

\newcommand{\ep}{1/\eps}

\newcommand{\one}{(\mathbf{1})}
\newcommand{\two}{(\mathbf{2})}
\newcommand{\otwo}{(\mathbf{12})}
\newcommand{\RV}{(\mathbf{RV})}
\newcommand{\LO}{{\tiny{\mbox{LO}}}}
\newcommand{\NLO}{{\tiny{\mbox{NLO}}}}
\newcommand{\NNLO}{{\tiny{\mbox{NNLO}}}}

\newcommand{\nnb}{\nonumber}

\newcommand{\beq}{\begin{eqnarray*}}
\newcommand{\eeq}{\end{eqnarray*}}

\newcommand{\Si}{{\bf S}_i}

\newcommand{\Cj}{{\bf C}_{\ij}}

\newcommand{\SCj}{{\bf SC}_{\ij k}}
\newcommand{\CSj}{{\bf CS}_{\ij k}}
\newcommand{\SCl}{{\bf SC}_{ikl}}

\newcommand{\CCk}{{\bf C}_{ik\!j}}

\newcommand{\CCl}{{\bf C}_{\ij kl}}
\newcommand{\SSk}{{\bf S}_{ik}}
\newcommand{\SSj}{{\bf S}_{\ij}}

\newcommand{\dphit}{\frac{d\Phi_{n+2}}{d\Phi_n}}
\newcommand{\dphio}{\frac{d\Phi_{n+2}}{d\Phi_{n+1}}}

\title{Towards analytic local sector subtraction at NNLO}

\ShortTitle{Towards analytic local sector subtraction at NNLO}

\author{Lorenzo Magnea,~~Ezio Maina,~~\speaker{Paolo Torrielli},~~Sandro Uccirati\\
Dipartimento di Fisica and Arnold-Regge Center, Universit\`a di Torino,\\
        and INFN, Sezione di Torino, Via P. Giuria 1, I-10125 Torino, Italy, \\
        E-mail: \email{magnea@to.infn.it, maina@to.infn.it, torriell@to.infn.it, uccirati@to.infn.it}}

\abstract{A new method for local subtraction at next-to-next-to-leading 
order in QCD is sketched, attempting to conjugate the minimal counterterm
structure arising from a sector partition of the radiation phase space with 
the simplifications following from analytic integration of the counterterms.}

\FullConference{13th International Symposium on Radiative Corrections\\
		24-29 September, 2017\\
		St. Gilgen, Austria}

\begin{document}

\section{Introduction}
\label{intro}

The next-to-next-to-leading perturbative order (NNLO) in QCD is rapidly 
becoming the new accuracy standard for fixed-order cross-section predictions 
at colliders. Calculations beyond leading order (LO) receive contributions from 
virtual and real radiation; when considered separately, these contributions 
generate infrared and collinear (IRC) singularities, that however cancel 
upon combination into physical cross sections\footnote{UV renormalisation 
and collinear factorisation are understood.}. Since the complexity of the processes 
under consideration requires numerical techniques to evaluate the relevant 
amplitudes, it becomes necessary to address the problem of getting rid of 
IRC singularities before the final numerical evaluation.

The subtraction technique achieves this goal systematically and with no 
approximations, by adding and subtracting to the  cross sections a set of 
local counterterms with the same singular behaviour as the real matrix 
elements in all unresolved corners of phase space. Upon analytic integration, 
these give rise to the same singularities as the virtual corrections. The 
universality of the IRC behaviour of gauge-theory amplitudes ensures the 
existence of such counterterms, and thus the applicability of a subtraction 
method.

At next-to-leading order (NLO) the problem has been solved two decades ago, 
and the main recipes developed in the literature are the FKS~\cite{Frixione:1995ms} 
and the CS~\cite{Catani:1996vz} methods. At NNLO, a number of subtraction 
methods have been proposed and employed to produce important phenomenological 
results, among them antenna subtraction \cite{GehrmannDeRidder:2005cm}, 
sector-improved residue subtraction \cite{Czakon:2010td,Boughezal:2011jf}, 
colourful subtraction \cite{Somogyi:2005xz}, $\mc E$-prescription
\cite{Frixione:2004is}, projection to Born
\cite{Cacciari:2015jma}\footnote{Slicing methods are also
  available at NNLO. The main ones are $q_T$ \cite{Catani:2007vq} and
  $N$-jettiness \cite{Boughezal:2015dva,{Gaunt:2015pea}}.}.
Still, the considerable increase in complexity in the proposed schemes in comparison
with the available NLO solutions motivates further investigation, especially 
considering that, for some of them, complexity implies forgoing desirable features 
such as locality or analyticity.

In view of trying to solve the NNLO problem in full generality and at minimal 
computational cost, and eventually in the hope of extending the procedure 
to yet higher orders, we believe it is worthwhile to re-examine some of the 
fundamental questions about the nature of the subtraction mechanisms, 
such as what are the simplest possible structures capable of achieving a 
local subtraction, how the available freedom in the definition of counterterms
can be fully exploited, and what are the ideas, among those successfully 
applied at NLO, that can be advantageously exported to the next order(s). 
In this contribution we present the preliminary results of this investigation
\footnote{For further developments and details see \cite{Magnea:2018hab}.}, 
for now limited to processes featuring only final-state massless QCD 
partons.

\section{NLO analysis}
\label{NLO}

At NLO, for a generic $2 \to n$ process, the differential cross section with 
respect to an IRC-safe observable $X$ can be schematically written as
\begin{eqnarray}
\label{eq:NLO_struct}
  (d \sig_\NLO - d \sig_\LO)/dX \, = \, \int d \Phi_n \, V \, \delta_n + 
  \int d \Phi_\npo \, R \, \delta_\npo \, ,
\end{eqnarray}
where $R$ and $V$ are the real and UV-renormalised virtual corrections, 
and $\delta_i \equiv \delta(X - X_i)$. $V$ features up to a double $\ep$ 
pole ($\eps$ being the dimensional regulator, $d = 4 - 2 \eps$), while $R$ 
is finite in $d = 4$, but features up to two singular limits in the radiation phase 
space. The subtraction procedure amounts to adding and subtracting a
counterterm $\int  d \tilde \Phi_\npo \, K \, \delta_n$, with $d \tilde \Phi_\npo 
\, K$ featuring the same phase-space singularities as $d \Phi_\npo \, R$, but 
at the same time being sufficiently simple to be analytically integrated in $d$ 
dimensions. Denoting the integrated counterterm with
\begin{equation}
\label{eq:int_count}
  I \, = \, \int \frac{d \tilde \Phi_\npo}{d \Phi_n} \, K \, ,
\end{equation}
the subtracted cross section becomes
\begin{equation}
  (d \sig_\NLO - d \sig_\LO)/dX \, = \, \int d \Phi_n (V + I) \, \delta_n + 
  \int \left( d \Phi_\npo \, R \, \delta_\npo - d \tilde\Phi_\npo \, K \, \delta_n \right) \ ,
\end{equation}
which is manifestly finite in $d = 4$ and integrable numerically.

\subsection{FKS subtraction}
\label{FKS}

The problem of finding and integrating the function $K$ is considerably simplified 
by introducing, as done by FKS, a partition of the radiation phase space in sectors, 
in each of which only up to two identified partons can give rise to IRC singularities. 
This is achieved by introducing sector functions $\mc W_{\ij}$ (with $\ij$ the singular 
pair, $j \neq i$), normalised through $\sum_{\ij}\mc W_{\ij} = 1$, which dampen all 
real-radiation singularities except the ones stemming from configurations where 
$i$ becomes soft ($\Si$ limit) and $\ij$ become collinear ($\Cj$ limit). Moreover 
one requires that the following properties be satisfied
\begin{equation}
\label{eq:sec_prop}
  \Si \sum_k \mc W_{ik} \, = \, 1 \, , \qquad \qquad
  \Cj \left( \mc W_{\ij} + \mc W_{ji} \right) \, = \, 1 \, ,
\end{equation}
implying that, by summing over the sectors whose functions do not vanish in the 
$\Si$ and $\Cj$ limits, the functions disappear. This feature is crucial for analytic 
counterterm integration: since the integrated counterterm is to be eventually 
combined in (\ref{eq:NLO_struct}) with the virtual contribution, which is not split 
into sectors, it is convenient to sum over sectors before analytic integration, 
thus getting rid of the explicit  (and potentially complicated) functional form of 
$\mc W_{\ij}$, by means of (\ref{eq:sec_prop}). The sectors are thus useful 
when combining the real correction with the counterterm into a finite quantity 
in $d=4$, but their presence must not complicate the analytic part of the 
computation.

Sectors however do not uniquely define the subtraction scheme: freedom is 
left in the parametrisation of the radiation phase space in each sector, in the 
kinematic mapping that allows to factorise exactly the Born result from the 
radiation phase space, so as to integrate the countertem only in the latter, 
and in the choice of the non-singular contributions to be included in the 
definition of the counterterm.

In FKS, the radiation phase space in each sector $\ij$ is parametrised 
independently, in terms of the rescaled energy $\xi_i = 2 E_i/\sqrt s$ of 
parton $i$ ($\Si = \lim_{~\xi_i \to 0}$), and of the cosine $y_{\ij} = \cos 
\theta_{\ij}$ of the angle between partons $\ij$ ($\Cj = \lim_{~y_{\ij} \to 
1}$)\footnote{A third variable, the azimuth $\phi_i$ of parton $i$ with 
respect to a given reference direction, is understood.}, where all quantities 
are defined in the center-of-mass frame of the collision, with energy 
$\sqrt s$. The kinematic mapping is defined once the sector is specified, 
by means of an appropriate common Lorentz boost of all particles but 
$i$, $j$. With this parametrisation, the counterterm in sector $\ij$ is 
defined as the collection of the singular terms in the Laurent expansion 
of the real correction around the IRC limits,
\begin{equation}
\label{eq:FKS_cnt}
  d \tilde \Phi_\npo^{(\ij)} \, K_{\ij} \, = \, \big( \Si + \Cj - \Si \Cj \big) 
  d \Phi_\npo \, R \, \mc W_{\ij} \, ,
\end{equation}
and the full counterterm is
\begin{equation}
\label{eq:FKS_cnt_full}
  d \tilde \Phi_\npo \, K \, = \, \sum_{\ij} d \tilde \Phi_\npo^{(\ij)} \, K_{\ij} \, .
\end{equation}
In (\ref{eq:FKS_cnt}), the ordering of the limits in the third term has been 
chosen arbitrarily, as the latter do commute ($\xi_i$ and $1 - y_{\ij}$ are 
allowed to tend to 0 independently).

\subsection{Bottlenecks of FKS in view of NNLO}
\label{bottle}

FKS defines a natural and compact subtraction scheme, however some 
of its features are not optimal towards analytical simplicity; all of them are 
fully manageable at NLO, owing to the straightforward structure of the 
relevant IRC kernels, but these seeds of complication may eventually 
hamper an analytic treatment of counterterms at the next perturbative orders.

As an example, by parametrising \emph{before} defining the counterterm, 
FKS looses some freedom in its analytic integration: the soft limit $\Si$ 
features an eikonal double sum $\sum_{kl}\frac{s_{kl}}{s_{ik}s_{il}}$ that 
results in
\begin{equation}
\label{eq:soft_FKS}
  \int \Si \, \sum_j d \Phi_\npo R \, \mc W_{\ij} \, \propto \, 
  \sum_{kl} \int d \Omega_i \frac{1 - \cos\theta_{kl}}{(1 - \cos\theta_{ki})
  (1 - \cos\theta_{il})} \, ,
\end{equation}
where $s_{ab} = 2 p_a \cdot p_b$, and $\Omega_i$ is the solid angle of 
parton $i$. This is not immediately trivial because the eikonal kernel involves 
invariants that do not belong to the sector for which the parametrisation 
has been devised, and the freedom in re-parametrising them is reduced 
after the soft variable $\xi_i$ has been pulled out in the limit (see right-hand 
side of (\ref{eq:soft_FKS})).

This difficulty is also partly related to the non-Lorentz-invariance of the FKS 
variables, which may represent a bottleneck at NNLO: the double-unresolved 
kernels \cite{Catani:1998nv,Catani:1999ss} are compact in terms of $s_{ab}$, 
but hardly  manageable analytically if parametrised with energies and angles.

Finally, the $d$-dimensional radiation phase space in the FKS parametrisation is
\begin{equation}
  \frac{d \Phi_{\npo}}{d \Phi_n} \, \propto \, d \xi_i \, dy_{\ij} 
  \left[ {2 - \xi_i (1 - y_{\ij})} \right]^{2\eps} \, ,
\end{equation}
which is immediately integrated only because the parenthesis trivialises in 
all IRC limits relevant to sector $\ij$; at NNLO it may be problematic to find 
a parametrisation with such a feature, that still respects the commutation 
properties of the composite limits (see comments below (\ref{eq:FKS_cnt_full})).

\subsection{Modified sector subtraction at NLO}
\label{modFKS}

The above bottlenecks can be alleviated by means of the following considerations. 
First, the singularities in sector $\ij$ are known once the identity of partons $i$ 
and $j$ is given, hence a local counterterm can be defined without referring to 
any specific parametrisation, by collecting the singular limits of the real-radiation 
matrix element, written in terms of dot products $s_{ab}$ of four-momenta. Second, it is not 
necessary that all contributions to the counterterm in a sector feature the 
same parametrisation or kinematic mapping: the latter can be chosen so 
as to maximally simplify the integration of the selected contribution.

The first of these considerations allows us to introduce $K_{\ij}$ through 
the following procedure.
\begin{itemize}
\setlength\itemsep{0mm}
\item Define the behaviour of (functions of) invariants in the singular limits:
\begin{eqnarray}
\label{eq:soft_prop}
  &&\mbox{soft}~i,~\Si:\quad s_{ia}/s_{ib}\to \mbox{constant} \, ,
  \quad s_{ia}/s_{bc}\to 0 \, ,\quad\forall\,a,b,c\ne i \, , \\
\label{eq:coll_prop}
  &&\mbox{collinear}~\ij,~\Cj:\quad s_{\ij}/s_{ab}\to 0,\quad s_{ia}/s_{\!\!ja}\to 
  \mbox{independent of $a$} \, ,\quad\forall\,ab\ne \ij \, .
\end{eqnarray}
\item Define $\Si R\,\mc W_{\ij}$ and $\Cj R\,\mc W_{\ij}$ as the most singular 
terms in the Laurent expansion of $R\,\mc W_{\ij}$ around the IRC limits, 
according to the scaling in (\ref{eq:soft_prop}) and (\ref{eq:coll_prop}). 
In particular
\begin{equation}
\label{eq:NLO_lim}
  \Si R~\propto~-\delta_{ig}\sum_{kl}\frac{s_{kl}}{s_{ki}s_{il}}B_{kl},
  \qquad \Cj R~\propto~ \frac1{s_{\ij}}P(z_{\ij})B \, ,
\end{equation}
where $\delta_{ig}$ forces the soft parton to be a gluon, $B$ and $B_{kl}$ 
are the Born and colour-linked Born squared matrix elements, $P$ is the 
relevant Altarelli-Parisi collinear kernel, and $z_{\ij}=s_{ir}/(s_{ir}+s_{jr})$, 
with arbitrary $r \neq \ij$. 
\item Define the counterterm in sector $\ij$ as
\begin{equation}
\label{eq:mod_cnt}
  d\Phi_\npo K_{\ij} \, = \, d\Phi_\npo \big(\Si+\Cj-\Si\Cj\big)R \,\mc 
  W_{\ij}~=~d\Phi_\npo \left[1-(1-\Si)(1-\Cj)\right]R \,\mc W_{\ij} \, .
\end{equation}
\end{itemize}

The order in which the $\Si$ and $\Cj$ operators appear in the composite 
limit is arbitrary. While in FKS the chosen parametrisation must explicitly 
realise such a commutation of limits, in order for composite residues to 
be defined, in this modified framework commutation naturally stems from 
fundamental properties of the soft and collinear limits, which are physically 
independent. Once the counterterm is defined as in (\ref{eq:mod_cnt}), a 
subsequent parametrisation of the latter in terms of non-independent 
variables is allowed, and does not spoil any of its properties.

Equations (\ref{eq:FKS_cnt}) and (\ref{eq:mod_cnt}) are structurally 
very similar and clearly share the same singular terms, showing that 
the modified scheme defines as minimal a local subtraction procedure 
as the original FKS; the two prescriptions differ by finite contributions, 
precisely those that make the counterterm in (\ref{eq:mod_cnt}) 
parametrisation-independent. Moreover, in (\ref{eq:mod_cnt}) the 
phase space associated with the counterterm is exact, namely the 
soft and collinear limits are applied only to matrix elements and sector 
functions. While this property is not crucial, and could immediately be 
lifted if required by computational convenience, it displays the enhanced 
flexibility of the modified scheme: as a parametrisation has not been 
chosen at this point yet, one has still the freedom to select one in 
which the phase space is trivial everywhere, without being compelled 
to evaluate the latter in the IRC limits.

The second of the above considerations allows to choose kinematic 
mappings and parametrisation independently of the sector. A particularly 
convenient choice of mapping is the one introduced by CS, where 
the $\npo$ real momenta $p_i$ are mapped on $n$ Born-like momenta 
$\bar p_j$ (the latter entering the computation as arguments of $B$ and 
$B_{kl}$ in (\ref{eq:NLO_lim})) through
\begin{eqnarray}
\label{eq:CSmap}
  &&\bar p_{c} = \frac1{1-y}p_c \, , \qquad \bar p_{[ab]}=p_a+p_b-\frac y{1-y}p_c \, ,
  \qquad \bar p_l=p_l,  \quad \forall \, l \ne a,b,c \, , \\
\label{eq:CSparam}
  && y = y_{abc}=\frac{s_{ab}}{s_{ab}+s_{ac}+s_{bc}} \, ,
  \qquad z=z_{abc} = \frac{s_{ac}}{s_{ac}+s_{bc}} \, .
\end{eqnarray}
In the hard-collinear counterterm in sector $\ij$, $(\Cj-\Si\Cj)R \,\mc W_{\ij}$ 
in (\ref{eq:mod_cnt}), labels are assigned as $a\!=\!i$, $b\!=\!j$, $c\!=\!r$, 
where $i$ and $j$ define the sector, while $r$ appears in the definition of 
$z_{\ij}$ in (\ref{eq:NLO_lim}). In the soft counterterm, $\Si R \,\mc W_{\ij}$, 
\emph{each term} of the sum over $kl$ \emph{is mapped differently}, with 
$a\!=\!i$, $b\!=\!k$, $c\!=\!l$. The phase space is parametrised in terms of 
variables $y_{abc}$ and $z_{abc}$ defined in (\ref{eq:CSparam}), with 
labels $abc$ assigned according to the relevant kinematic mapping, as 
just described. In particular, in a given sector, not all contributions to the 
counterterm are parametrised in the same way, the latter indeed being 
the feature that complicates the integration of the soft counterterm in FKS.

Each term in the eikonal double sum is now straightforwardly integrated:
\begin{eqnarray}
\label{eq:soft_mod}
  \int \frac{d\Phi_\npo}{d\Phi_n}\frac{s_{kl}}{s_{ki}s_{il}} & \propto & (\bar p_{[ik]}
  \cdot \bar p_l)^{-  \eps}\int_0^1 dz\int_0^1 dy\,\Big[y(1-y)^2z(1-z)\Big]^{-\eps} \,
  \frac{(1-y)(1-z)}{yz},\nnb\\
  &=&(\bar p_{[ik]}\cdot \bar p_l)^{-\eps}B(-\eps,2-\eps)B(-\eps,2-2\eps) \, ,
\end{eqnarray}
where $z=z_{ikl}$, $y=y_{ikl}$, and $B$ is the Euler beta function, a result 
valid to all orders in $\eps$.

The modified sector subtraction outlined in this section successfully works 
at NLO, as the integrated counterterm can be shown to analytically reproduce 
all virtual poles. The method, to some extent, bridges the FKS and CS approaches, 
retaining the strengths of both, in particular sector partition and minimal 
counterterm structure from FKS, and Lorentz invariance and phase-space 
mappings from CS. We believe this approach to be more easily exportable 
to NNLO, since it achieves the maximal possible simplification as far as analytic 
integration is concerned.

\section{NNLO analysis}
\label{NNLO}

At NNLO, the differential cross section with respect to IRC-safe observable $X$ is
\begin{equation}
\label{eq:NNLO_struct}
  (d\sig_\NNLO-d\sig_\NLO)/dX \, = \, \int d\Phi_n\,VV\,\delta_n+
  \int d\Phi_\npo\,RV\,\delta_\npo+\int d\Phi_\npt\,RR\,\delta_\npt \, ,
\end{equation}
where $RR$, $VV$, $RV$, are the double-real and UV renormalised double-virtual 
and real-virtual corrections. $VV$ features up to a quadruple $\ep$ pole, $RR$ is 
finite in $d=4$, but features up to four phase-space singularities, and $RV$ has 
up to a double $\ep$ pole and diverges doubly in the radiation phase space. 
The subtraction procedure amounts to adding and subtracting $\int  d \tilde 
\Phi_\npt\,\big[K^{\one}\,\delta_\npo+(K^{\otwo}+K^{\two})\,\delta_n\big]$, as well as $\int  
d\tilde\Phi_\npo\,K^{\RV}\,\delta_n$, where $K^{\one}$ and $(K^{\otwo}+K^{\two})$ are 
the single- and double-unresolved counterterms, containing all singularities 
of $RR$ in the limits where one or two partons become unresolved\footnote{In
the case of the double-unresolved counterterm, $K^{\two}$ collects all `homogeneous'
double-unresolved configurations, namely the ones where two partons become
unresolved with the same scaling, while $K^{\otwo}$ contains all `hierarchical'
double-unresolved configurations, namely the ones where two partons become
unresolved in a strongly-ordered manner.}, while
$K^{\RV}$ is the real-virtual counterterm, featuring the same phase-space 
singularities as $RV$. Denoting the corresponding integrated counterterms with
\begin{equation}
\label{eq:int_count_NNLO}
  I^{(\bf p)}\, = \, \int\frac{d\tilde\Phi_\npt}{d\Phi_{n\!+\!2\!-\!p}}K^{(\bf p)},
  \qquad I^{\otwo}\, = \, \int\frac{d\tilde\Phi_\npt}{d\Phi_{n\!+\!1}}K^{\otwo},
  \qquad I^{\RV}=\int\frac{d\tilde\Phi_\npo}{d\Phi_n}K^{\RV}, \qquad p=1,2 \, ,
\end{equation}
the subtracted cross section becomes
\begin{eqnarray}
  \hspace{-5mm}
  \label{eq:nnlostruct}
  (d\sig_\NNLO-d\sig_\NLO)/dX & = & \int d\Phi_n (VV+I^{\two}+I^{\RV})\,
  \delta_n\nnb\\
  &&+\int[(d\Phi_\npo\, RV+d\tilde\Phi_\npo\, I^{\one})\,\delta_\npo-
  d\tilde\Phi_\npo\,(K^{\RV}-I^{\otwo})\delta_n]\nnb\\
  &&+\int [d\Phi_\npt \,RR\,\delta_\npt-d\tilde\Phi_\npt\,K^{\one}\,
  \delta_\npo-d\tilde\Phi_\npt(K^{\otwo}+K^{\two})\delta_n] \, .
\end{eqnarray}
$I^{\one}$ features the same $\ep$ poles as $RV$, $I^{\otwo}$ features the same
$\ep$ poles as $K^{\RV}$, while the sum $I^{\two} + I^{\RV}$ has the same $\ep$
poles as $VV$, ensuring all contributions are finite in $d=4$ and integrable
numerically.

\subsection{Modified sector subtraction at NNLO}
\label{modNNLO}

In order to define an analytic subtraction procedure at NNLO, it is convenient 
to divide the phase space in sectors, in each of which only up to four identified 
partons can give rise to IRC singularities. Each sector function $\mc W_{abcd}$ 
($abcd$ being the singular combinations, $b\ne a$, $c\ne a$, $d\ne a,c$), 
normalised through $\sum_{abcd} \mc W_{abcd} = 1$, dampens all double-real 
singularities, except a single-soft and a single-collinear ($\Si$, $\Cj$ in sectors 
$\ij k\!j$, $\ij\!jk$, and $\ij kl$), a double-soft ($\SSk$ in sectors ${\ij k\!j}$ and 
${\ij kl}$, $\SSj$ in sector ${\ij\!jk}$), a double-collinear ($\CCk$ in sectors 
${\ij k\!j}$ and ${\ij\!jk}$, $\CCl$ in sector ${\ij kl}$), and a soft-collinear
($\SCj$ in sector ${\ij k\!j}$, $\SCj$ and $\CSj$ in sector ${\ij\!jk}$, $\SCl$ and
$\CSj$ in sector $\ij kl$). To clarify: in  configuration ${\bf S}_{ab}$, partons
$ab$ are all soft; in ${\bf C}_{abc}$, partons $abc$ are all collinear, while in
${\bf C}_{abcd}$ the four partons become collinear in pairs; in ${\bf SC}_{abc}$,
hierarchically, $a$ becomes soft, then $bc$ become collinear; in ${\bf CS}_{abc}$,
$ab$ become collinear, then $c$ becomes soft. ${\bf SC}_{abc}$ and ${\bf CS}_{bca}$ are
the same projector when acting on matrix elements, but not on sector functions.
Roughly speaking, sectors $\ij k\!j$ and $\ij\!jk$ select singularities
associated with splitting $a\to \ij k$, while  sector $\ij kl$ is
associated with independent splittings $a_1\to \ij$, and $a_2\to kl$.

The next step, in analogy with (\ref{eq:sec_prop}), is to enforce the constraint 
that sector functions disappear upon summation over the sectors whose 
functions do not vanish in double-unresolved limits. This requirement, 
crucial for the analytic integration of $K^{\two}$, reads
\begin{eqnarray}
\label{eq:sec_prop_NNLO}
  \SSk \sum_{d\neq i,k}\Big(\sum_{b\neq i}\mc W_{ibkd} +
  \sum_{b\neq k}\mc W_{kbid}\Big) \, = \, 1 \, ,
  \qquad \CCk \sum_{abc\,\in\,\scriptsize{\mbox{perm}}~\ij k} 
  \Big(\mc W_{abbc}+\mc W_{abcb}\Big) \, = \, 1 \, ,
\end{eqnarray}
and analogously for projectors ${\bf C}_{ijkl}$, ${\bf SC}_{abc}$, and
${\bf CS}_{abc}$, where perm $ik$ = $ik,$ $ki$,
while perm $\ij k$ = $\ij k$, $ik\!j$, $jik$, $jki$, 
$k\ij$, $k\!ji$. At NNLO, however, one more constraint has to be satisfied: 
as $RV$ is split into NLO-type sectors $\mc W_{\ij}$, since it has single-real 
kinematics, $I^{\one}$ must feature the same $\ep$ poles as $RV$, 
\emph{NLO sector by NLO sector}, in order for $d \Phi_\npo\, RV + 
d \tilde\Phi_\npo\, I^{\one}$ to be finite for each $\ij$ independently. 
This is achieved by requiring the NNLO sector functions to factorise 
the NLO ones in the single-unresolved limits, as
\begin{eqnarray}
  \Cj \mc W_{\ij k\!j}&\sim&\bar \mc W_{k[\ij]}\,\Cj \mc 
  W_{\ij},\qquad \Si \mc W_{\ij k\!j}~\sim~\bar \mc W_{k\!j}\,\Si \mc W_{\ij} \, , \\
  \Cj \mc W_{\ij\!jk}&\sim&\bar \mc W_{[\ij]k}\,\Cj \mc 
  W_{\ij},\qquad \Si \mc W_{\ij\!jk}~\sim~\bar \mc W_{jk}\,\Si \mc W_{\ij} \, , \\
  \Cj \mc W_{\ij kl}&\sim&\bar \mc W_{kl}\,\Cj \mc W_{\ij},\qquad \,\,\,\,\,\,
  \Si \mc W_{\ij kl}~\sim~\bar \mc W_{kl} \, \Si \mc W_{\ij} \, ,
\end{eqnarray}
where the bars denote kinematic mappings analogous to the ones described 
in (\ref{eq:CSmap}).

The local counterterms are defined in analogy with (\ref{eq:mod_cnt}), as
\begin{eqnarray}
\label{eq:cnt_NNLO_loc1}
  K^{\one}_{\ij k\!j}+K^{\otwo}_{\ij k\!j}+K^{\two}_{\ij k\!j} & = & \left[1-(1-\Si)(1-\Cj)(1-\SSk)
  (1-\CCk)(1-\SCj)\right]RR\,\mc W_{\ij k\!j} \, , \\
%
  K^{\one}_{\ij\!jk}+K^{\otwo}_{\ij\!jk}+K^{\two}_{\ij\!jk} & = & \left[1-(1-\Si)(1-\Cj)(1-\SSj)
  (1-\CCk)(1-\SCj)(1-\CSj)\right]RR\,\mc W_{\ij\!jk} \, , \nnb\\
%
  K^{\one}_{\ij kl}+K^{\otwo}_{\ij kl}+K^{\two}_{\ij kl} & = & \left[1-(1-\Si)(1-\Cj)(1-\SSk)
  (1-\CCl)(1-\SCl)(1-\CSj)\right]RR\,\mc W_{\ij kl} \, . \nnb
\end{eqnarray}
The kernels $\SSk$, $\CCk$, $\SCj$, and $\CSj$ are universal, and have been computed 
in \cite{Catani:1998nv,Catani:1999ss,Berends:1988zn}. The order of the various 
operators in the composite limits is arbitrary, as all limits commute.

Equations (\ref{eq:cnt_NNLO_loc1}) are appropriate 
to define local counterterms, but redundant: in particular, $RR$ can 
feature at most four singularities, hence not all operators that appear 
in those equations are `primary', namely carry independent information on the 
singularity structure of $RR$. These redundancies are readily eliminated by 
considering the idempotence of projection operators: for instance, once 
${\bf{SC}}_{iab}$ has been applied to a given quantity, further acting on 
it with $\Si$ does not produce any effect, and similarly for the action of $\CSj$ 
after $\Cj$ has been applied. The same is true for $\CCl$ and $\Cj$ when acting on matrix elements, but not on sector functions. One thus has $\Si\SCj=\SCj$, $\Cj\CSj=\CSj$, which implies
\begin{equation}
(1-\Si)\SCj ~=~ (1-\Si)\SCl ~=~ (1-\Cj)\CCl ~=~ 0.
\end{equation}
As a consequence, all soft-collinear double-unresolved limits, $\SCj$, $\SCl$, 
and $\CSj$, completely disappear from the sum $K^{\otwo}+ K^{\two}$
(see also \cite{Caola:2017dug} about the redundancy of the soft-collinear 
limit)\footnote{The integrals $I^{\two}$ and $I^{\otwo}$ have to be evaluated
separately, see Eq.~(\ref{eq:nnlostruct}), hence the kernels $\bf SC$ and $\bf CS$
do contribute in that case, even if they would cancel in the sum.}. Equations (\ref{eq:cnt_NNLO_loc1}) finally can be simplified to (with $T = \ij\!jk,\, \ij k\!j,\, \ij kl$)
\begin{eqnarray}
  K^{\one}_T & = & [\Si+\Cj(1-\Si)]RR\,\mc W_T , \\
  K^{\two}_{\ij\!jk}  & = & [\SSj+\CCk(1-\SSj)+\SCj(1-\SSj)(1-\CCk)]RR\,\mc W_{\ij\!jk} \, , \nnb\\
  K^{\two}_{\ij k\!j} & = & [\SSk+\CCk(1-\SSk)+(\SCj+\CSj)(1-\SSk)(1-\CCk)]RR\,\mc W_{\ij k\!j} \, , \nnb\\
  K^{\two}_{\ij kl} & = & [\SSk+\CCl(1-\SSk)+(\SCl+\CSj)(1-\SSk)(1-\CCl)]RR\,\mc W_{\ij kl} \, , \nnb\\
  K^{\otwo}_{\ij\!jk} & = & -\{[\Si+\Cj(1-\Si)][\SSj+\CCk(1-\SSj)]+\SCj(1-\SSj)(1-\CCk)\}RR\,\mc W_{\ij\!jk} \, , \nnb\\
  K^{\otwo}_{\ij k\!j} & = & -\{[\Si+\Cj(1-\Si)][\SSk+\CCk(1-\SSk)]+(\SCj+\CSj)(1-\SSk)(1-\CCk)\}RR\,\mc W_{\ij k\!j} \, , \nnb\\
  K^{\otwo}_{\ij kl} & = & -\{[\Si+\Cj(1-\Si)][\SSk+\CCl(1-\SSk)]+(\SCl+\CSj)(1-\SSk)(1-\CCl)\}RR\,\mc W_{\ij kl} \, , \nnb
\end{eqnarray}
where we have separated the counterterms according to the type of singularities
they feature: single-unresolved in $K^{\one}$, pure double-unresolved in $K^{\two}$, overlaps
of single- and double-unresolved projectors in $K^{\otwo}$.

\subsection{Counterterm integration}
\label{coint}

The integration of the double-unresolved counterterm proceeds from the 
definitions of $K^{\two}$ and crucially benefits from the defining properties in (\ref{eq:sec_prop_NNLO}), which allow to completely get rid of sector functions before
analytical integration. Indeed one gets
\begin{eqnarray}
  I^{\two}
  & = & \int \dphit\,\sum_i\Big[\sum_{j>i}\SSk+\sum_{j>i}\sum_{k>j}\CCk(1-\SSk-\SSj-{\bf S}_{\!jk})\nnb\\
  & & \hspace{5mm} + \sum_{j>i}\sum_{k>i\atop k\neq j}\sum_{l>k \atop l\neq j} \CCl(1-\SSk-{\bf S}_{\!jk}-{\bf S}_{il}-{\bf S}_{jl}) \nnb\\
  & & \hspace{5mm} + \sum_{j \neq i}\sum_{k\neq i \atop k > j} \SCj(1-\SSk-\SSj)(1-\CCk-\sum_{l\neq i,j,k}{\bf C}_{il\!jk}) \nnb\\
  & & \hspace{5mm} + \sum_{j > i}\sum_{k\neq i,j} \CSj(1-\SSk-{\bf S}_{\!jk})(1-\CCk-\sum_{l\neq i,j,k}\CCl) \Big] RR \, .
\end{eqnarray}
Since the sector functions have disappeared from the integrand, and only 
the singular kernels are left over, the integration can be managed analytically 
in $d = 4 - 2 \eps$ dimensions. As an explicit example of the computation, 
consider the case in which a $q\bar q$ pair becomes soft, which leads to 
the  double-soft kernel \cite{Catani:1999ss}
\begin{equation}
\label{eq:SS_NNLO}
  \SSk \, RR \, \propto \, (\as\,\mu^{2\eps})^2\,\,T_R \!\!\sum_{l,m=1}^n B_{lm} \,
  \frac{s_{il}s_{km}+s_{im}s_{kl}-s_{ik}s_{lm}}{s_{ik}^2(s_{il}+s_{kl})(s_{im}+s_{km})} \, ,
\end{equation}
with $\mu$ the renormalisation scale. Each term in the double sum in 
(\ref{eq:SS_NNLO}) is associated with a different CS mapping, as was 
the case for the soft term at NLO, in order to optimise the parametrisation 
for each addend separately. Denoting with $z'$, $y'$ the CS variables 
relevant to dipole $(ik,l)$, and with $z$, $y$ those relevant to dipole 
$([ik]l,m)$, the double-soft integrand for $l\neq m$ (for $l=m$ the 
result is trivial) after azimuthal integration is
\begin{eqnarray}
  \frac{s_{il}s_{km}+s_{im}s_{kl}-s_{ik}s_{lm}}{s_{ik}^2(s_{il}+s_{kl})(s_{im}+s_{km})} 
  & \propto & \frac{z'(1-z')}{y^2 y'^2 }\frac{z-y'(1-z) }{z+y'(1-z)} \, ,
\end{eqnarray}
to be integrated with the measure $\int_0^1 dy' dz' dy \, dz \, \left[
y' (1-y')^2 y^2 (1-y)^2 z(1-z) \right]^{- \eps} (1 - y') y (1 - y)$. The final 
result for $n = 2$ Born-level particles, integrated over the Born phase 
space, and with prefactors reinstated, reads
\begin{eqnarray}
  \int d\Phi_{\npt}\,\SSk\, RR & = &\sigma_\LO\left(\frac{\as}{2\pi}\right)^2
  T_RC_F\left(\frac{\mu^2}{s}\right)^{2\eps}\nnb\\
  & & \hspace{-25mm}\times\,\left[-\frac{1}{3\eps^3}-\frac{17}{9\eps^2}+
  \frac{1}{\eps}\left(\frac{7}{18}\pi^2-\frac{232}{27}\right)+
  \frac{38}{9}\zeta_3+\frac{131}{54}\pi^2-\frac{2948}{81}
  \right]+\mc O(\eps) \, .
\end{eqnarray}
The double-collinear limit relevant for a splitting $q \to qq'\bar q'$ is mapped 
and parametrised in a similar fashion, resulting in an integral of comparable 
complexity. One finds
\begin{eqnarray}
  \int d\Phi_{\npt}\, \CCk \,RR &=&\sigma_{\LO}\left(\frac{\as}{2\pi}\right)^2T_RC_F  
  \left(\frac{\mu^2}{s}\right)^{2\eps}\nnb\\
  && \hspace{-25mm} \times \, \left[-\frac{1}{3\eps^3}-\frac{31}{18\eps^2}+
  \frac{1}{\eps}\left(\frac{1}{2}\pi^2-\frac{889}{108}\right)+
  \frac{80}{9}\zeta_3+\frac{31}{12}\pi^2-\frac{23941}{648}
  \right]+\mc O(\eps) \, .
\end{eqnarray}
It has to be noted that double-unresolved limits involving gluons are more 
complicated than the one detailed here, but still manageable analytically.

\subsection{Proof-of-concept example}
\label{proco}

Considering the $T_RC_F$ contribution to the NNLO total cross section 
for $e^+e^-\!\to q(1)\bar q(2)$, the double-real process is $e^+e^-\!\to q(1)
\bar q(2) q'(3) \bar q'(4)$. All relevant matrix elements can be found 
in \cite{Ellis:1980wv,Hamberg:1990np,GehrmannDeRidder:2004tv}. 
Limits ${\bf S}_{34}$, ${\bf C}_{134}$, ${\bf C}_{234}$, and ${\bf C}_{34}$ 
are non-zero, and the integrated counterterms read
\begin{eqnarray}
\int\!\!d\Phi_n\,I^{\two}\!\!
&=&\!
\int d\Phi_\npt\,\big[
 {\bf S}_{34}
+{\bf C}_{134}(1-{\bf S}_{34})
+{\bf C}_{234}(1-{\bf S}_{34})
\big]RR
\nnb\\
&=&\!
\sigma_\LO\!\left(\frac{\as}{2\pi}\right)^{\!\!\!2}\!T_R\,C_F\!\!
\left(\!\frac{\mu^2}{s}\!\right)^{\!\!\!\!2\eps}\!\left[
-\frac{1}{3\eps^3}
-\frac{14}{9\eps^2}
+\frac{1}{\eps}\!\left(\!\frac{11}{18}\pi^2-\frac{425}{54}\!\right)\!
+\frac{122}{9}\zeta_3
+\frac{74}{27}\pi^2
-\frac{12149}{324}
\right]\!\!,
\nnb\\
I^{\one}&=&I^{\one}_{12}+I^{\one}_{1[34]}+I^{\one}_{2[34]},\nnb\\
I^{\one}_{hq}&=&\,\bar{\mc W}_{hq}\int \dphio\,{\bf C}_{34}\,RR
 \, = \, -\frac{\as}{2\pi}\left(\frac{\mu^2}{s}\right)^\eps \frac23 T_R
 \Big[\frac1\eps-\ln\frac{\bar s_{[34]r}}{s}+\frac83\Big]R\bar \mc W_{hq}+\mc O(\eps).
\end{eqnarray}
The structure of $\bar{\mc W}$ functions appearing in the addends of $I^{\one}$ 
is precisely the one of the subtracted real-virtual contribution, split into NLO 
sectors.
The sums $RV_{hq}^{\rm fin} \equiv RV \bar{\mc W}_{hq}+I^{\one}_{hq}$ are finite in $d=4$
\begin{eqnarray}
RV_{hq}^{\rm fin}
&=&-\frac\as{2\pi}\frac23 T_R\left(\ln\frac{\mu^2}{\bar s_{r[34]}}+\frac{8}3\right) R\,\bar{\mc W}_{hq} + \mc O(\eps),
\end{eqnarray}
with $r=1$ or 2 when $hq=12$, while $r=3-h$ in the other cases. Analogously, the sum $K^{\RV}_{hq}-I^{\otwo}_{hq}$ is finite in $d=4$, and reads
\begin{eqnarray}
K^{\RV}_{hq}-I^{\otwo}_{hq}
&=&-\frac\as{2\pi}\frac23 T_R\left(\ln\frac{\mu^2}{\bar s_{r[34]}}+\frac{8}3\right)[{\bf S}_h+{\bf C}_{hq}(1-{\bf S}_h)] R\,\bar{\mc W}_{hq} + \mc O(\eps).
\end{eqnarray}
The integrated real-virtual counterterm is
\begin{eqnarray}
\int d\Phi_n\,I^{\RV}
&=&
\sum_{\ij}\int d\Phi_\npo\,K^{\RV}_{\ij}
\nnb\\
&=&
\frac\as{2\pi}\frac23\frac{T_R}\eps \int d\Phi_\npo\,\big[
{\bf S}_{[34]} + {\bf C}_{1[34]}(1-{\bf S}_{[34]}) + {\bf C}_{2[34]}(1-{\bf S}_{[34]})
\big]R
\nnb\\
&=&
\sigma_\LO\left(\frac\as{2\pi}\right)^{\!2}T_R\,C_F\!
\left(\frac{\mu^2}{s}\right)^{\!\!\eps}\left[
 \frac4{3\eps^3}
+\frac2{\eps^2}
+\frac1\eps\left(-\frac79\pi^2+\frac{20}3\right)
-\frac{100}9\zeta_3
-\frac{7}{6}\pi^2
+20
\right].
\nnb
\end{eqnarray}
Collecting all contributions, for instance setting $\mu=0.35\sqrt s$, one gets
\begin{eqnarray}
\label{eq:VV_subt}
  \int d\Phi_n\,(VV+I^{\two}+I^{\RV}) & = &\sigma_\LO
  \left(\frac\as{2\pi}\right)^2T_RC_F 
  \left(\frac83\zeta_3-\frac19\pi^2-\frac{44}9-\frac43\ln\frac{\mu^2}s\right) \nnb \\
  & = & \sigma_\LO\left(\frac\as{2\pi}\right)^2T_RC_F \times 0.01949914 \, , \\
\label{eq:RV_subt}
  \int d\Phi_\npo\,(RV+I^{\one}-K^{\RV}) & = & \sigma_\LO
  \left(\frac\as{2\pi}\right)^2T_RC_F \times(-0.90635 \pm 0.00011) \, , \\
\label{eq:RR_subt}
  \int d\Phi_\npt\,(RR-K^{\one}-K^{\two}) & = & \sigma_\LO
  \left(\frac\as{2\pi}\right)^2T_RC_F\times(+2.29491 \pm 0.00038) \, ,
\end{eqnarray}
where (\ref{eq:VV_subt}) is a fully analytic result, in (\ref{eq:RV_subt}) 
the cancellation of $\ep$ poles is analytic, and the remaining finite integral 
is numerical, while (\ref{eq:RR_subt}) is fully numerical.

By summing (\ref{eq:VV_subt}) to (\ref{eq:RR_subt}), the NNLO correction 
obtained with the subtraction method is
\begin{equation}
  \frac1{\left(\frac\as{2\pi}\right)^2T_RC_F}
  \frac{\sig_{\NNLO}-\sig_{\NLO}}{\sig_{\LO}}=1.40806 \pm 0.00040 \, ,
\end{equation}
to be compared with the analytic result
$-11/2+4\zeta_3\!-\ln(\mu^2/s)=1.40787186$.
The plot below shows that the renormalisation-scale dependence is also  
correctly reproduced.
\begin{center}
  \includegraphics[width=0.85\textwidth]{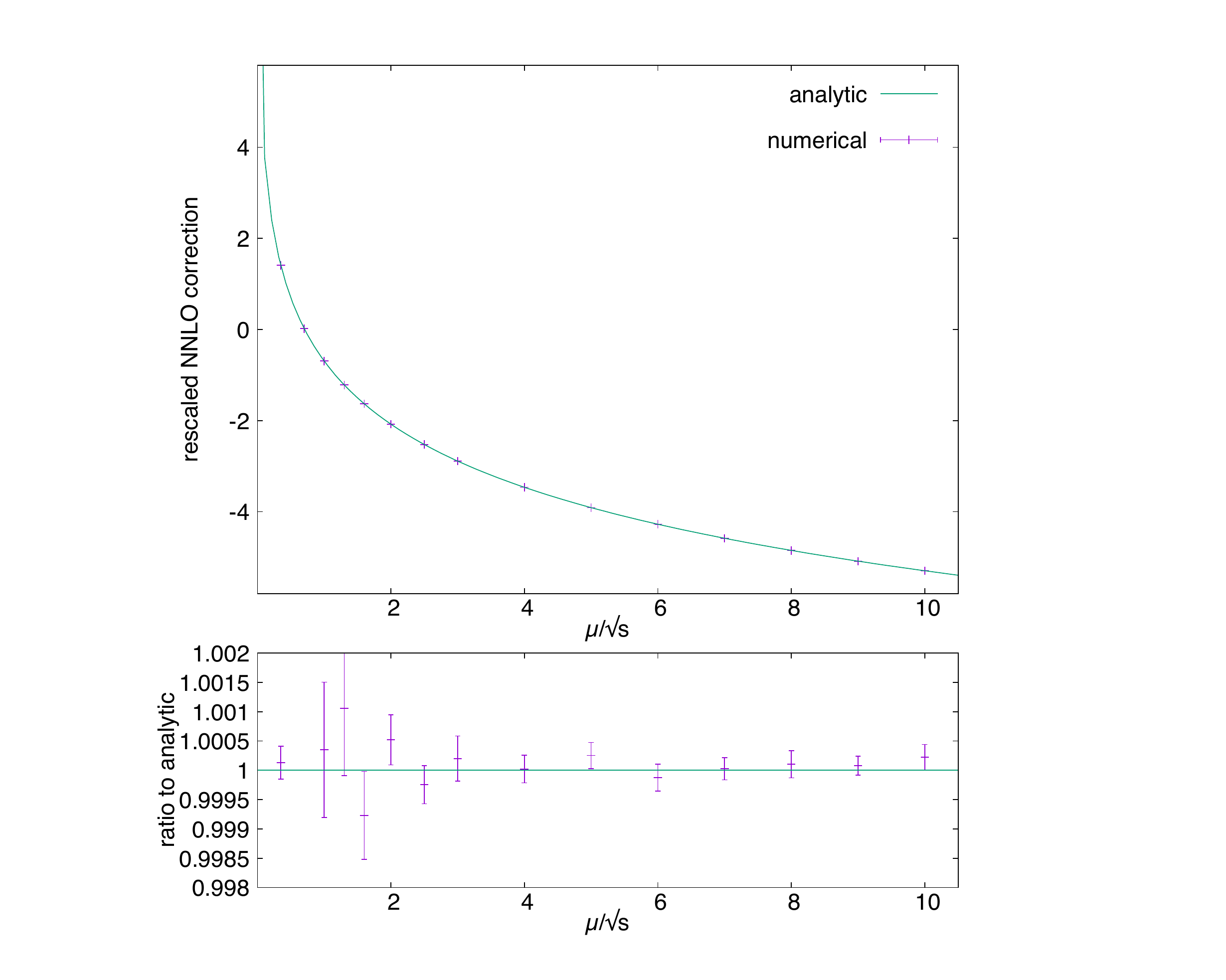}
\end{center}

\section{Conclusions}
We have presented the theoretical basis of a new method for NNLO local 
sector subtraction, aiming at minimality in the definition of the counterterms, 
and analyticity in their integration. The method has been presented in the 
NLO case, and applied to a simplified case at NNLO, displaying the 
expected properties. Generalisations to the complete NNLO case are ongoing.

\section*{Acknowledgements}
The work of PT has received funding from the European Union Seventh 
Framework programme for research and innovation under the Marie Curie 
grant agreement N. 609402-2020 researchers: Train to Move (T2M).


\begin{thebibliography}{99}
\bibitem{Frixione:1995ms}
  S.~Frixione, Z.~Kunszt and A.~Signer,
  Nucl.\ Phys.\ B {\bf 467} (1996) 399
  [hep-ph/9512328].
  S.~Frixione,
  Nucl.\ Phys.\ B {\bf 507} (1997) 295
  [hep-ph/9706545].

\bibitem{Catani:1996vz}
  S.~Catani and M.~H.~Seymour,
  Nucl.\ Phys.\ B {\bf 485} (1997) 291
  [hep-ph/9605323].
  S.~Catani, S.~Dittmaier, M.~H.~Seymour and Z.~Trocsanyi,
  Nucl.\ Phys.\ B {\bf 627} (2002) 189
  [hep-ph/0201036].
  

\bibitem{GehrmannDeRidder:2005cm}
  A.~Gehrmann-De Ridder, T.~Gehrmann and E.~W.~N.~Glover,
  JHEP {\bf 0509} (2005) 056
  [hep-ph/0505111].
  A.~Daleo, T.~Gehrmann and D.~Maitre,
  JHEP {\bf 0704} (2007) 016
  [hep-ph/0612257].
T.~Gehrmann, these proceedings.
    
\bibitem{Czakon:2010td}
  M.~Czakon,
  Phys.\ Lett.\ B {\bf 693} (2010) 259
  [arXiv:1005.0274 [hep-ph]].
  M.~Czakon,
  Nucl.\ Phys.\ B {\bf 849} (2011) 250
  [arXiv:1101.0642 [hep-ph]].
A.~Behring, these proceedings.

\bibitem{Boughezal:2011jf}
  R.~Boughezal, K.~Melnikov and F.~Petriello,
  Phys.\ Rev.\ D {\bf 85} (2012) 034025
  [arXiv:1111.7041 [hep-ph]].
 R.~Roentsch, these proceedings. 

\bibitem{Somogyi:2005xz}
  G.~Somogyi, Z.~Trocsanyi and V.~Del Duca,
  JHEP {\bf 0506} (2005) 024
  [hep-ph/0502226].
  G.~Somogyi, Z.~Trocsanyi and V.~Del Duca,
  JHEP {\bf 0701} (2007) 070
  [hep-ph/0609042].
A.~Kardos, these proceedings. 

\bibitem{Frixione:2004is}
  S.~Frixione and M.~Grazzini,
  JHEP {\bf 0506} (2005) 010
  [hep-ph/0411399].


\bibitem{Cacciari:2015jma}
  M.~Cacciari, F.~A.~Dreyer, A.~Karlberg, G.~P.~Salam and G.~Zanderighi,
  Phys.\ Rev.\ Lett.\  {\bf 115} (2015) no.8,  082002
  [arXiv:1506.02660 [hep-ph]].
M.~Cacciari, these proceedings.

\bibitem{Catani:2007vq}
  S.~Catani and M.~Grazzini,
  Phys.\ Rev.\ Lett.\  {\bf 98} (2007) 222002
  [hep-ph/0703012].


\bibitem{Boughezal:2015dva}
  R.~Boughezal, C.~Focke, X.~Liu and F.~Petriello,
  Phys.\ Rev.\ Lett.\  {\bf 115} (2015) no.6,  062002
  [arXiv:1504.02131 [hep-ph]].


\bibitem{Gaunt:2015pea}
  J.~Gaunt, M.~Stahlhofen, F.~J.~Tackmann and J.~R.~Walsh,
  JHEP {\bf 1509} (2015) 058
  doi:10.1007/JHEP09(2015)058
  [arXiv:1505.04794 [hep-ph]].


\bibitem{Magnea:2018hab}
  L.~Magnea, E.~Maina, G.~Pelliccioli, C.~Signorile-Signorile, P.~Torrielli and S.~Uccirati,
  arXiv:1806.09570 [hep-ph].
  

\bibitem{Catani:1998nv}
  S.~Catani and M.~Grazzini,
  Phys.\ Lett.\ B {\bf 446} (1999) 143
  [hep-ph/9810389].
   
\bibitem{Catani:1999ss}
  S.~Catani and M.~Grazzini,
  Nucl.\ Phys.\ B {\bf 570} (2000) 287
  [hep-ph/9908523].

\bibitem{Berends:1988zn}
  F.~A.~Berends and W.~T.~Giele,
  Nucl.\ Phys.\ B {\bf 313} (1989) 595.

\bibitem{Caola:2017dug}
  F.~Caola, K.~Melnikov and R.~Röntsch,
  Eur.\ Phys.\ J.\ C {\bf 77} (2017) no.4,  248
  [arXiv:1702.01352 [hep-ph]].

\bibitem{Ellis:1980wv}
  R.~K.~Ellis, D.~A.~Ross and A.~E.~Terrano,
  Nucl.\ Phys.\ B {\bf 178} (1981) 421.
  
\bibitem{Hamberg:1990np}
  R.~Hamberg, W.~L.~van Neerven and T.~Matsuura,
  Nucl.\ Phys.\ B {\bf 359} (1991) 343
   Erratum: [Nucl.\ Phys.\ B {\bf 644} (2002) 403].

\bibitem{GehrmannDeRidder:2004tv}
  A.~Gehrmann-De Ridder, T.~Gehrmann and E.~W.~N.~Glover,
  Nucl.\ Phys.\ B {\bf 691} (2004) 195
  [hep-ph/0403057].
\end{thebibliography}
\end{document}